\documentclass[twocolumn,aps,superscriptaddress,showpacs,nofootinbib,floatfix]{revtex4-1}

\usepackage{graphicx}
\usepackage{epsfig}
\usepackage{bm}
\usepackage{amssymb}
\usepackage{amsmath}
\usepackage{slashed}

\usepackage[normalem]{ulem}
\usepackage{color}

\renewcommand{\sout}{\bgroup \color{red} \ULdepth=-.5ex \ULset}

\begin{document}

\title{Role of $\Sigma^*(1385)$ in $\Lambda$ hyperon polarization in relativistic heavy ion collisions}

\author{Haesom Sung}
\email{ioussom@gmail.com}
\affiliation{Institute of Physics, Academia Sinica, Taipei 11529, Taiwan}

\author{Che Ming Ko}\email{ko@comp.tamu.edu}
\affiliation{Cyclotron Institute and Department of Physics and Astronomy, Texas $A\&M$ University, College Station, TX 77843, USA}

\author{Su Houng Lee}\email{suhoung@yonsei.ac.kr}
\affiliation{Department of Physics and Institute of Physics and Applied Physics, Yonsei University, Seoul 03722, Korea}

\begin{abstract}
The effect of $\Sigma^*(1385)$ baryon resonance on the time evolution of the $\Lambda$ hyperon polarization in hadronic matter is studied using a spin-dependent kinetic approach.  This approach explicitly includes the production of the $\Sigma^*$ resonance from the $\Lambda-\pi$ and $\Sigma(1192)-\pi$ scatterings as well as its decay into $\Lambda+\pi$ or $\Sigma+\pi$, all including explicitly their spin dependence.  The resulting coupled kinetic equations governing the time evolution of $\Lambda$, $\Sigma$ and $\Sigma^*$ numbers and polarizations are solved for Au-Au collisions at $\sqrt{s_{NN}}=7.7$ GeV and 20-50\% centrality, using initial values determined by their thermal yields and the thermal vorticity in the hadronic matter at chemical freeze-out temperature.  As the hadronic matter expands and cools, the $\Lambda$ polarization is found to increase slightly during early times and then decreases very slowly afterwards, while the $\Sigma$ polarization remains nearly constant and the $\Sigma^*$ polarization continuously decreases.  Including feed-down contributions to the $\Lambda$ polarization from the decays of partially polarized $\Sigma^0$, $\Sigma^*$, and $\Xi(1322)$ hyperons, where the $\Xi$ polarization is obtained by solving coupled kinetic equations for the $\Xi$ and $\Xi^*(1532)$ system, the resulting $\Lambda$ polarization becomes smaller and decreases over time. In both cases, the time variation of the $\Lambda$ polarization is, however, sufficiently small to support the assumption of an early freeze-out of $\Lambda$ spin degree of freedom in relativistic heavy ion collisions.
\end{abstract}

\maketitle

\section{Introduction}

Since the observation by the STAR Collaboration in 2017 that the $\Lambda$ hyperon produced in heavy ion collisions at the Relativistic Heavy Ion Collider (RHIC) are partially polarized along the direction perpendicular to the reaction plane~\cite{STAR:2017ckg}, numerous theoretical studies have been carried out to understand the origin of this phenomenon~\cite{Becattini:2020ngo}.   Most of these studies are based on hydrodynamic models~\cite{Karpenko2017,Fu:2021pok,Becattini:2021iol,Palermo:2024tza} or the course-grained transport model~\cite{Li:2017slc}, which assume thermal equilibrium between the $\Lambda$ spin and the vortical fluid, and calculate the global $\Lambda$ polarization from the thermal vorticity at the chemical freeze-out temperature.  
Extensions of these studies have incorporated the effects of thermal shear in the fluid to address the measured $\Lambda$ local polarizations~\cite{Niida:2018hfw}, particularly its component along the beam direction, as functions of the azimuthal angle in the transverse plane of the collision~\cite{Liu:2021uhn,Becattini:2021suc,Yi:2021ryh}. 

Besides equilibrium models, the $\Lambda$ polarization has also been studied using non-equilibrium framework such as the chiral kinetic approach~\cite{Sun:2016mvh,Sun:2018bjl} and the covariant angular-momentum-conserved chiral transport model~\cite{Liu:2019krs}. In these studies, both vorticity and shear in the quark matter affect the dynamics of quarks and antiquarks through their equations of motion and scatterings. The polarization of the $\Lambda$ hyperon is then determined via the coalescence of polarized strange quarks with polarized up and down quarks during hadronization of the quark matter~\cite{Sun:2017xhx}.  

In all these theoretical studies, the $\Lambda$ polarization is calculated at the chemical freeze-out temperature and compared with the measured values obtained after the kinetic freeze-out of the hadronic matter.  A recent study based on the assumption of thermal equilibrium between the $\Lambda$ spin and the thermal vorticity and shear shows that the $\Lambda$ polarization would decrease during the hadronic evolution due to the reduction of vorticity and shear with decreasing temperature and increasing volume of the hadronic matter~\cite{Sun:2021nsg}. The fact that experimental data are consistent with the $\Lambda$ polarization at the hadronization temperature suggests that the $\Lambda$ spin freezes out, or decouples, early from the hadronic matter. 

As proposed in Ref.~\cite{Ko:2023eyb}, this early freeze-out would occur if $\Lambda-\pi$ scattering proceeds through the spin 3/2, positive parity $\Sigma^*(1385)$ resonance, since the ratio of $\Lambda$ spin non-flip to flip probabilities in such scattering is 3.5. This explanation has been supported  by detailed theoretical calculations of the $\Lambda$ polarization relaxation time in hot hadronic matter, which is found to be about 4-7 fm/$c$ ~\cite{Hidaka:2023oze}, and by the kinetic approach to the time evolution of $\Lambda$ polarization during the hadronic stage of relativistic heavy ion collisions, which show a decreases of about 7-12\%~\cite{Sung:2024vyc}.   However, these studies are based on the approximation of instantaneous $\Lambda-\pi$ resonance scattering via the $\Sigma^*$ resonance, and neglect feed-down contributions to the $\Lambda$ polarization from other hadrons.  Since the experimentally measured $\Lambda$ polarization includes not only primary $\Lambda$ hyperons but also those from the decays of resonances~\cite{Becattini:2016gvu,Xia:2019fjf} -- particularly the strong decay $\Sigma^*\to\Lambda+\pi$, the electromagnetic decay $\Sigma^0(1192)\to\Lambda+\gamma$, and the weak decays $\Xi^0(1315)\to\Lambda+\pi^0$ and $\Xi^-(1322)\to\Lambda+\pi^-$ -- these contributions must be considered. 

In the present study, we go beyond the approximation used in the kinetic approach of Ref.~\cite{Sung:2024vyc} by treating the $\Sigma^*$ resonance dynamically via the processes $\Lambda+\pi\leftrightarrow \Sigma^*$, i.e., the production of $\Sigma^*$ from $\Lambda-\pi$ scattering and its decay back to $\Lambda$ and $\pi$. In addition, we consider feed-down contributions to the $\Lambda$ polarization from the decays of $\Sigma^0$ and $\Sigma^*$ as well as from $\Xi^0$ and $\Xi^-$, with the latter obtained by solving coupled kinetic equations for the numbers and polarizations of $\Xi$ and $\Xi^*(1532)$ hyperons during the hadronic stage of relativistic heavy ion collisions. 

This paper is organized as follows. Section~\ref{cross} introduces the cross sections for the $\Lambda+\pi\to\Sigma^*$ and $\Sigma+\pi\to\Sigma^*$ reactions, together with the decay widths for their inverse processes. Section~\ref{thermal} presents calculations of their thermal averages in a hot hadronic matter.  In Sec.~\ref{kinetic}, we derive the kinetic equations for the numbers and polarizations of $\Lambda$, $\Sigma$, and $\Sigma^*$ hyperons. Section~\ref{results} provides solutions to these equation and discusses the resulting time evolution.  Section~\ref{feed-down} describes the feed-down contributions to $\Lambda$ polarization, including those  from $\Xi$ and $\Xi^*$ based on their coupled kinetic equations.  Finally, conclusions are presented in Sec.~\ref{conclusions}.

\section{$\Lambda+\pi\leftrightarrow\Sigma^*\leftrightarrow\Sigma+\pi$ processes}\label{cross}

\begin{figure}[h]
    \centering  \includegraphics[width=0.9\linewidth]{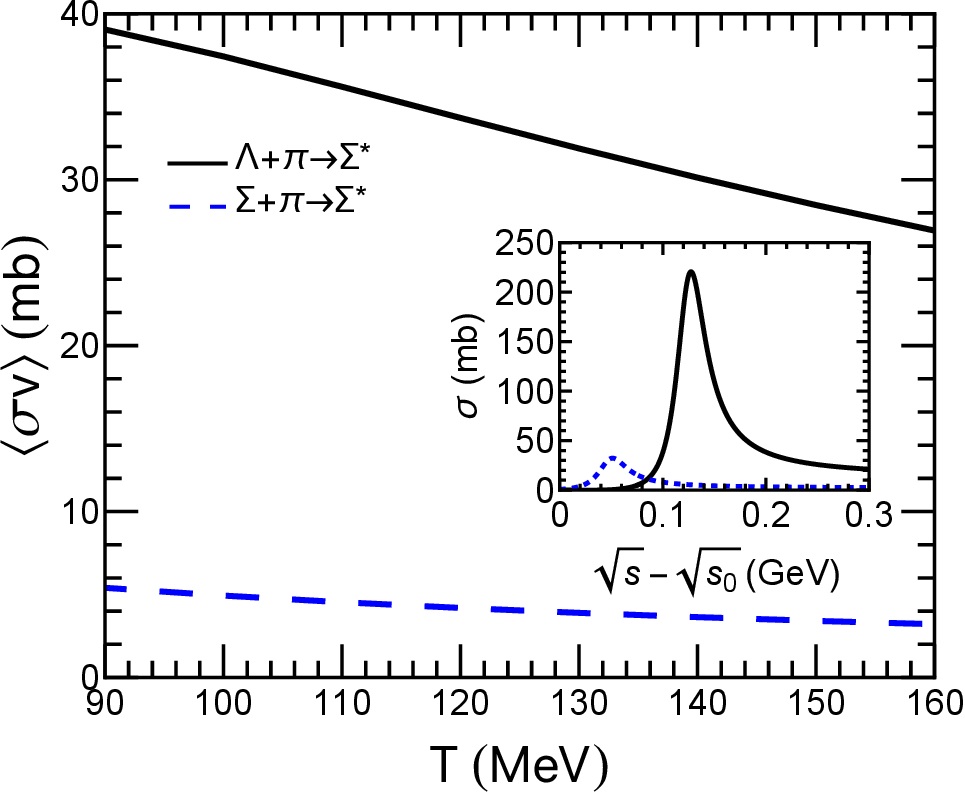}
     \caption{Temperature dependence of the thermal averages of the spin-averaged $\Lambda+\pi\rightarrow \Sigma^*$ (solid line) and $\Sigma+\pi\rightarrow \Sigma^*$ (dashed line) cross sections shown in the inset.}
     \label{fig:thermal}
\end{figure}

For the $\Lambda+\pi\rightarrow \Sigma^*$ and $\Sigma+\pi\rightarrow \Sigma^*$ scatterings at center-of-mass energy $\sqrt{s}$, their spin-averaged cross sections take the usual Breit-Wigner form,  
\begin{eqnarray}
&&\sigma_{Y+\pi\to\Sigma^*}(s)=\frac{2s_{\Sigma^*}+1}{(2s_{Y}+1)(2s_{\pi}+1)}\frac{4\pi}{k^2}\notag\\
&&\hspace{1.2cm}\times\frac{B_{\Sigma^*\to Y+\pi}s\Gamma_{\Sigma^*}^2(s)}{(s-m_{\Sigma^*}^2)^2+s\Gamma_{\Sigma^*}^2(s)}, \quad Y=\Lambda,\Sigma.
    \label{eq:resonance}
\end{eqnarray}
In the above, $s_\pi=0$, $s_Y=1/2$ and $s_{\Sigma^*}=3/2$ are the spins of pion, hyperon $Y$ (with $Y=\Lambda, \Sigma$), and the $\Sigma^*$ resonance, respectively.  The variable $k$ denotes the magnitude of the momenta of $Y$ and $\pi$ in their center-of-mass frame, which is determined by the center-of-mass energy $\sqrt{s}$ and the masses of the pion ($m_\pi=140$ MeV) and hyperon $Y$ ($m_Y=1116$ MeV for $\Lambda$ and 1192 MeV for $\Sigma$) through the relation: 
\begin{equation}\label{momentum}
    k=\sqrt{\frac{[s-(m_{Y}+m_{\pi})^2][s-(m_{Y}-m_{\pi})^2]}{4s}}.
\end{equation}
The energy-dependent width $\Gamma_{\Sigma^*}(s)$ in Eq.(\ref{eq:resonance}) represents the total width of the $\Sigma^*$ resonance and is given by the sum of its partial widths, i.e., $\Gamma_{\Sigma^*}(\sqrt{s})=\sum_{Y=\Lambda, \Sigma}\Gamma_{\Sigma^*\to Y+\pi}(\sqrt{s})$. Each partial width $\Gamma_{\Sigma^*\to Y+\pi}(\sqrt{s})$ is related to its value at the resonance mass $\sqrt{s}=m_{\Sigma^*}$ by $\Gamma_{\Sigma^*\to Y+\pi}(\sqrt{s})=\Gamma_{\Sigma^*\to Y+\pi}(m_{\Sigma^*})(k/k_0)^3$, where $k_0$ and $k$ are the center-of mass momenta of hyperon $Y$ and pion at energy $\sqrt{s}=m_{\Sigma^*}$ and at $\sqrt{s}$, respectively. Empirically, one has $\Gamma_{\Sigma^*}(m_{\Sigma^*})=38$ MeV, with  branching ratios $B_{\Sigma^*\to\Lambda+\pi}=0.88$ and $B_{\Sigma^*\to\Sigma+\pi}=0.12$.   The spin-averaged cross sections for the reactions $\Lambda+\pi\rightarrow \Sigma^*$ and $\Sigma+\pi\rightarrow \Sigma^*$, calculated from Eq.(\ref{eq:resonance}), are shown in the inset of Fig.~\ref{fig:thermal} by solid and dashed lines, respectively.   As expected, both cross sections peak at $\sqrt{s}$ equal to $m_{\Sigma^*}$ above the respective threshold energy $\sqrt{s_0}=m_Y+m_\pi$.

To study the effect of hadronic scatterings on $\Lambda$ polarization in relativistic heavy ion collisions, spin-dependent cross sections are required.  These are related to the spin-averaged cross sections $\sigma_{Y+\pi\to\Sigma^*}$ in Eq.(\ref{eq:resonance}) via the Clebsch-Gordan coefficients $\langle j_{Y} m_{Y} l_{\pi}m_{\pi} | j_{\Sigma^*} {m}_{\Sigma^*}\rangle$ that couple the hyperon spin state $|j_{Y},m_{Y}\rangle$ and pion orbital angular momentum state $|l_{\pi},m_{\pi}\rangle$ to the $\Sigma^*$ spin state $| j_{\Sigma^*} {m}_{\Sigma^*}\rangle$. Specifically. the spin-dependent cross sections are:
$\sigma_{Y_{\pm 1/2}+\pi\rightarrow \Sigma^*_{\pm 3/2}}=\sigma_{Y+\pi\to\Sigma^*}/2$, $\sigma_{{Y_{\pm 1/2}}+\pi\rightarrow \Sigma^*_{\pm 1/2}}=\sigma_{Y+\pi\to\Sigma^*}/3$, and $\sigma_{Y_{\pm 1/2}+\pi\rightarrow \Sigma^*_{{\mp 1/2}}}=\sigma_{Y+\pi\to\Sigma^*}/6$.  Similarly, the spin-dependent partial decay widths of the $\Sigma^*$ resonance are related to the spin-averaged partial decay width $\Gamma_{\Sigma^*\to Y+\pi}$ as follows: $\Gamma_{\Sigma^{*}_{\pm 3/2}\rightarrow Y_{\pm1/2}+\pi} = \Gamma_{\Sigma^*\to Y+\pi}$, $\Gamma_{\Sigma^{*}_{\pm 1/2}\rightarrow Y_{\pm1/2}+\pi} =2\Gamma_{\Sigma^*\to Y+\pi}/3$ and $\Gamma_{\Sigma^{*}_{\pm 1/2}\rightarrow Y_{\mp1/2}+\pi} =\Gamma_{\Sigma^*\to Y+\pi}/3$.

\section{thermally averaged cross sections}\label{thermal}

For $\Lambda$ and $\Sigma$ hyperons scattering with pions in a thermally equilibrated hadronic matter at temperature $T$, their scattering effects are described by the thermally averaged cross sections $\langle\sigma_{\Lambda+\pi\to\Sigma^*} v\rangle$ and $\langle\sigma_{\Sigma+\pi\to\Sigma^*} v\rangle$, given by
\begin{eqnarray}\label{eq:AvgCrs}
&&\langle\sigma_{Y+\pi\to\Sigma^*} v\rangle =\frac{\int d^{3}{\bf p}_{Y}d^{3}{\bf p}_{\pi}f_{Y}({\bf p}_{Y})f_{\pi}({\bf p}_{\pi})\sigma_{Y+\pi\to\Sigma^*} v}{\int d^{3}{\bf p}_{Y}d^{3}{\bf p}_{\pi}f_{Y}({\bf p}_Y)f_{\pi}({\bf p}_{\pi})}\notag\\
&&\hspace{0.2in}=\frac{T^4}{4m_Y^2K_2(m_Y/T)m_\pi^2K_2(m_\pi/T)}\notag\\
&&\hspace{0.2in}\times\int_{z_0}^\infty dz K_1(z)\sigma_{Y+\pi\to\Sigma^*}(z^2T^2)\notag\\
&&\hspace{0.2in}\times[z^2-(m_Y+m_\pi)^2/T^2][z^2-(m_Y-m_\pi)^2/T^2].
\end{eqnarray}
In the above, $f_{Y}({\bf p}_{Y})$ is the Boltzmann momentum distribution, given by $f_Y({\bf p}_{Y})=e^{-(\sqrt{m_Y^2+{\bf p}_Y^2}-\mu_Y)/T}$, where the chemical potential $\mu_Y=B_{Y}\mu_{B}+S_{Y}\mu_{S}$ is expressed  in terms of the baryon number $B_Y$ and the strangeness number $S_Y$ of the hyperon, along with the baryon chemical potential $\mu_B$ and strangeness chemical potential $\mu_S$ of the hadronic matter.   The relative velocity between $Y$ and $\pi$ in Eq.(\ref{eq:AvgCrs}) is given by $v=\sqrt{(p_Y\cdot p_\pi)^2-m_Y^2 m_\pi^2}/(E_Y E_\pi)$, where $p_{Y,\pi}=(E_{Y,\pi}, {\bf p}_{Y,\pi})$ are their four momenta. The functions $K_1(z)$ and $K_2(m_i/T)$ are the modified Bessel functions of the first kind.

The thermal averages of the spin-averaged $Y-\pi$ scattering cross sections as functions of temperature are shown in Fig.~\ref{eq:resonance} by solid and dashed lines, respectively.  Although the scattering cross sections shown in the inset exhibit peak structures, their thermal averages are smooth functions of temperature that decrease with increasing $T$.  In particular, the thermally averaged $\Lambda-\pi$ cross section is about a factor of six larger than that of the $\Sigma-\pi$ cross section, owing to the larger decay branching ratio of the $\Sigma^*$ resonance to $\Lambda+\pi$ compared to $\Sigma+\pi$.

\section{Kinetic equations for $\Lambda$, $\Sigma$, and $\Sigma^*$ numbers and polarizations}\label{kinetic}

Assuming that pions are in thermal equilibrium within an expanding hadronic matter, as in Ref.~\cite{Sung:2024vyc}, the time evolution of the numbers $N_{\Lambda_m}$, $N_{\Sigma_m}$, and $N_{\Sigma^*_m}$ -- corresponding to $\Lambda$, $\Sigma$, and $\Sigma^*$ hyperons in spin states $m_{\Lambda}$, $m_\Sigma$, and $m_{\Sigma^*}$, respectively, -- is governed by a set of coupled kinetic equations.  Each equation includes a gain term and a loss term, arising from either the formation reactions $\Lambda_{m}+\pi\to\Sigma^*_{m^\prime}$ and $\Sigma_{m}+\pi\to\Sigma^*_{m^\prime}$, or the decay processes $\Sigma^*_{m}\to\Lambda_{m^\prime}+\pi$ and $\Sigma^*_{m}\to\Sigma_{m^\prime}+\pi$.  

Using the spin-dependent $Y+\pi\to\Sigma^*$ cross section in Sect.~\ref{cross}, the kinetic equations for $N_{\Lambda_m}$, $N_{\Sigma_m}$, and $N_{\Sigma^*_m}$ can be expressed in terms of the thermally averaged cross section $\langle\sigma_{Y+\pi\to \Sigma^*}v\rangle$ from Eq.(\ref{eq:AvgCrs}), and the similarly defined thermal averages of the $\Sigma^*$ decay widths, $\langle\Gamma_{\Sigma^*\to Y+\pi}\rangle$, as  
\begin{widetext}
\begin{eqnarray}\label{kinetic1}
&&\frac{dN_{Y_{1/2}
}}{d\tau} = \frac{1}{3}\langle\Gamma_{\Sigma^*\to Y+\pi}\rangle\left(3 N_{\Sigma^{*}_{3/2}}+2N_{\Sigma^{*}_{1/2}}+N_{\Sigma^{*}_{-1/2}} \right)
- \langle\sigma_{Y+\pi\to\Sigma^*} v\rangle z_{\pi}^{(T)}n_{\pi}^{(T)}N_{Y_{1/2}},\quad Y=\Lambda,\Sigma\notag \\ 
&&   \frac{dN_{Y_{-1/2}}}{d\tau}=\frac{1}{3}\langle\Gamma_{\Sigma^*\to Y+\pi}\rangle \left(N_{\Sigma^{*}_{1/2}}+2N_{\Sigma^{*}_{-1/2}}+3N_{\Sigma^{*}_{-3/2}} \right)- \langle\sigma_{Y+\pi\to\Sigma^*} v\rangle z_{\pi}^{(T)}n_{\pi}^{(T)}N_{Y_{-1/2}},\quad Y=\Lambda,\Sigma\notag \\
&&\frac{dN_{\Sigma^*_{3/2}}}{d\tau} = \frac{1}{2}\sum_{Y=\Lambda,\Sigma}\left[\langle\sigma_{Y+\pi\to\Sigma^*} v\rangle z_{\pi}^{(T)}n_{\pi}^{(T)}  N_{Y_{1/2}}-\langle \Gamma_{\Sigma^*\to Y+\pi}\rangle N_{\Sigma^{*}_{3/2}}\right], \notag \\
&&    \frac{dN_{\Sigma^*_{1/2}}}{d\tau} = \frac{1}{6}\sum_{Y=\Lambda,\Sigma}\left[\langle\sigma_{Y+\pi\to\Sigma^*} v\rangle z_{\pi}^{(T)}n_{\pi}^{(T)}(2N_{Y_{1/2}}+N_{Y_{-1/2}})-\langle \Gamma_{\Sigma^*\to Y+\pi}\rangle N_{\Sigma^*_{1/2}}\right],  \notag \\
&&    \frac{dN_{\Sigma^*_{-1/2}}}{d\tau} = \frac{1}{6} \sum_{Y=\Lambda,\Sigma}\left[\langle\sigma_{Y+\pi\to\Sigma^*} 
v\rangle z_{\pi}^{(T)}n_{\pi}^{(T)}(N_{Y_{1/2}}+2N_{Y_{-1/2}})-\langle \Gamma_{\Sigma^*\to Y+\pi}\rangle N_{\Sigma^*_{-1/2}}\right],\notag \\
&&    \frac{dN_{\Sigma^*_{-3/2}}}{d\tau} = \frac{1}{2} \sum_{Y=\Lambda,\Sigma}\left[\langle\sigma_{Y+\pi\to\Sigma^*} v\rangle z_{\pi}^{(T)}n_{\pi}^{(T)}N_{Y_{-1/2}}-\langle \Gamma_{\Sigma^*\to Y+\pi}\rangle N_{\Sigma^{*}_{-3/2}}\right].
\end{eqnarray}
\end{widetext}
In the above equations, $n_\pi^{(T)}$ denotes the thermally equilibrated pion density, given by 
\begin{eqnarray}
n_{\pi}^{(T)}=\frac{g_\pi}{(2\pi)^3}\int d^3{\bf p}\frac{1}{e^{\sqrt{m_\pi^2+{\bf p}^2}/T}-1}, 
\end{eqnarray}
where $g_\pi=3$ is the pion isospin degeneracy factor.  The pion fugacity $z_\pi^{(T)}$ in Eq.(\ref{kinetic1}) accounts for the increasing pion number due to the decays of resonances as the hadronic matter expands and cools.  

The kinetic equations in Eq.(\ref{kinetic1}) describing the time evolution of $\Lambda$, $\Sigma$, and $\Sigma^*$ in different spin states can be  reformulated as coupled kinetic equations for their total numbers, 
\begin{eqnarray}
&&N_Y=\sum_{m=-1/2, 1/2}N_{Y_m}, \quad Y=\Lambda, \Sigma\notag\\
&&N_{\Sigma^*}=\sum_{m=-3/2,-1/2,1/2,3/2}N_{\Sigma^*_m},
\end{eqnarray} 
and polarizations~\cite{Becattini:2016gvu},
\begin{eqnarray}\label{polariztion}
&&P_{Y}=\frac{N_{Y_{1/2}}-N_{Y_{-1/2}}}{N_{Y_{1/2}}+N_{Y_{-1/2}}},\quad Y=\Lambda, \Sigma\notag\\
&&P_{\Sigma^*}=\frac{N_{{\Sigma^*}_{3/2}}+ \frac{1}{3}N_{{\Sigma^*}_{1/2}}-\frac{1}{3}N_{{\Sigma^*}_{-1/2}}-N_{{\Sigma^*}_{-3/2}}}{N_{{\Sigma^*}_{3/2}}+N_{{\Sigma^*}_{1/2}}+N_{{\Sigma^*}_{-1/2}}+N_{{\Sigma^*}_{-3/2}}}, \notag\\
\end{eqnarray}
by taking appropriate combinations of the kinetic equations in Eq.(\ref{kinetic1}).  The resulting coupled kinetic equations are given by,
\begin{eqnarray}\label{kinetic2}
&&\frac{dN_{Y}}{d\tau} = \langle\Gamma_{\Sigma^*\to Y+\pi}\rangle N_{\Sigma^{*}}- \langle\sigma_{Y+\pi\to\Sigma^*} v\rangle z_{\pi}^{(T)}n_{\pi}^{(T)}N_{Y},\notag\\
&&\hspace{6cm}Y=\Lambda,\Sigma\notag \\ 
&&\frac{dN_{\Sigma^*}}{d\tau} = \sum_{Y=\Lambda,\Sigma}\left[\langle\sigma_{Y+\pi\to\Sigma^*} v\rangle z_{\pi}^{(T)}n_{\pi}^{(T)}N_{Y}\right.\notag\\
&&\hspace{1.3cm}-\left.\langle \Gamma_{\Sigma^*\to Y+\pi}\rangle N_{\Sigma^{*}}\right], \notag \\
&&\frac{dP_Y}{d\tau}=\langle\Gamma_{\Sigma^*\to Y+\pi}\rangle\frac{N_{\Sigma^*}}{N_Y}(P_{\Sigma^*}-P_Y),\quad Y=\Lambda,\Sigma\notag \\ 
&&\frac{dP_{\Sigma^*}}{d\tau}=z_\pi^{(T)}n_\pi^{(T)}\sum_{Y=\Lambda,\Sigma}\langle\sigma_{Y+\pi\to\Sigma^*} v\rangle\frac{N_Y}{N_{\Sigma^*}}
\notag\\
&&\hspace{1.3cm}\times\left(\frac{5}{9}P_Y-P_{\Sigma^*}\right).
\end{eqnarray}

Since the right-hand side of the kinetic equations in Eq.(\ref{kinetic2}) vanish when hyperons reach their thermally equilibrated values, given by $z_Y^{(T)}N_{Y}^{(T)}$, with $z_Y^{(T)}$ being the hyperon fugacity, terms involving $\langle\Gamma_{\Sigma^*\to Y+\pi}\rangle$ can be related to $\langle\sigma_{Y+\pi\to \Sigma^*} v\rangle$ through detailed balance, i.e.,  
\begin{eqnarray}\label{DB}
&&\langle\Gamma_{\Sigma^*\to Y+\pi}\rangle =\frac{\langle\sigma_{Y+\pi\to\Sigma^*} v\rangle z_{\pi}^{(T)}n_{\pi}^{(T)}z_Y^{(T)}N_{Y}^{(T)}}{z_{\Sigma^*}^{(T)}N_{\Sigma^*}^{(T)}}\notag\\
&&\hspace{1.2cm}=\frac{\langle\sigma_{Y+\pi\to\Sigma^*} v\rangle n_{\pi}^{(T)}N_{Y}^{(T)}}{N_{\Sigma^*}^{(T)}}.
\end{eqnarray}
In the above, we have used the relation $z_{\Sigma^*}^{(T)}=z_Y^{(T)} z_\pi^{(T)}$, which holds for thermally and chemically equilibrated $\pi$, $Y$ and $\Sigma^*$ numbers, to obtain the final expression. 

The quantities $N_Y^{(T)}$ and $N_{\Sigma^*}^{(T)}$ in Eq.(\ref{DB}) are given by
\begin{eqnarray}\label{equilibrium}
&&N_{Y}^{(T)}=\frac{g_YV}{(2\pi)^3}\int d^3{\bf p}e^{-(\sqrt{m_Y^2+{\bf p}^2}-\mu_Y)/T}\notag\\
&&\hspace{0.8cm}\approx\frac{g_YV}{2\pi^2}m_Y^2TK_2(m_Y/T)e^{\mu_Y/T},\quad Y=\Lambda, \Sigma\notag\\
&&N_{\Sigma^*_m}^{(T)}=\frac{A_{\Sigma^*}}{\pi}\int_{s_0}^\infty ds\frac{\sqrt{s}\Gamma_{\Sigma^*}(\sqrt{s})}{(s-m_{\Sigma^*}
^2)^2+s\Gamma_{\Sigma^*}^2(\sqrt{s})}\notag\\
&&\hspace{1.2cm}\times\frac{g_{\Sigma^*}V}{(2\pi)^3}\int d^3{\bf p}e^{-(\sqrt{s+{\bf p}^2}-\mu_{\Sigma^*})/T},
\end{eqnarray}
where $g_Y$ is the spin and isospin degeneracy of hyperon $Y$, i.e., $g_\Lambda=2$, $g_\Sigma=6$, and $g_{\Sigma^*}=12$, and $V$ denotes the volume of the hadronic matter.  Because of its finite width, the number of $\Sigma^*$ resonance is evaluated by including a mass distribution of the Breit-Wigner form, similar to that in its production cross section from $Y-\pi$ scattering in Eq.(\ref{eq:resonance}), with a normalization constant $A_{\Sigma^*}=0.69$ to ensure the integral of the mass distribution over $s$ from $s_0=(m_\Lambda+m_\pi)^2$ to $\infty$ to have a value of unity.   

Substituting Eq.(\ref{DB}) in Eq.(\ref{kinetic2}) leads to 
\begin{eqnarray}\label{kinetic3}
&&\frac{dN_{Y}}{d\tau} = \langle\sigma_{Y+\pi\to\Sigma^*} v\rangle z_{\pi}^{(T)}n_{\pi}^{(T)}\left(\frac{N_Y^{(T)}}{z_\pi^{(T)}N_{\Sigma^*}^{(T)}} N_{\Sigma^*}-N_Y\right),\notag\\  
&&\hspace{6cm}Y=\Lambda,\Sigma\notag \\ 
 &&\frac{dN_{\Sigma^*}}{d\tau}=\sum_{Y=\Lambda, \Sigma}\langle\sigma_{Y+\pi\to\Sigma^*} v\rangle z_{\pi}^{(T)}n_{\pi}^{(T)}\notag\\
&&\hspace{1.3cm}\times\left(N_Y-\frac{N_Y^{(T)}}{z_\pi^{(T)}N_{\Sigma^*}^{(T)}}N_{\Sigma^*}\right),\notag\\
&&\frac{dP_Y}{d\tau}=\langle\sigma_{Y+\pi\to\Sigma^*} v\rangle n_{\pi}^{(T)}\frac{N_Y^{(T)}N_{\Sigma^*}}{N_{\Sigma^*}^{(T)}N_Y }(P_{\Sigma^*}-P_{Y}),\notag\\
&&\hspace{6cm}Y=\Lambda,\Sigma\notag \\ 
&&\frac{dP_{\Sigma^*}}{d\tau}=\sum_{Y=\Lambda, \Sigma}\langle\sigma_{Y+\pi\to\Sigma^*} v\rangle z_{\pi}^{(T)}n_{\pi}^{(T)}\notag\\&&\hspace{1.3cm}\times\frac{N_Y}{N_{\Sigma^*}}\left(\frac{5}{9}P_Y-P_{\Sigma^*}\right).
\end{eqnarray}
These kinetic equations show that the total number of $\Lambda$, $\Sigma$ and $\Sigma^*$ remains constant in time, i..e, $\frac{d(N_\Lambda+ N_\Sigma+N_{\Sigma^*})}{d\tau}=0$. Whether the $\Lambda$ and $\Sigma$ polarizations increases or decreases with time depends on the signs of  $P_{\Sigma^*}-P_\Lambda$ and $P_{\Sigma^*}-P_\Sigma$, respectively.  In contrast, the time evolution of the $\Sigma^*$  polarization is governed by the combinations $5P_\Lambda/9-P_{\Sigma^*}$ and $5P_\Sigma/9-P_{\Sigma^*}$.

\section{results}\label{results}
 
In this section, we apply the kinetic equations given in Eq.(\ref{kinetic3}) to study the effect of hadronic scatterings on the $\Lambda$ polarization in Au-Au collisions at $\sqrt{s_{NN}}=7.7$ GeV and 20-50\% centrality~\cite{STAR:2017ckg}.   For the hadronization temperature and volume of the produced quark-gluon plasma, as well as the baryon and strangeness chemical potentials of the resulting hadronic matter, we use the values extracted by the STAR Collaboration from statistical model fits to measured particle yields in Au+Au collisions at $\sqrt{s_{NN}}=7.7$ GeV and 30-40\% centrality~\cite{STAR:2017sal}: $T_C=146$ MeV, $V_C=240$ fm$^3$, $\mu_{B}=376$ MeV, and $\mu_S = 88$ MeV.  

For the kinetic freeze-out temperature, we adopt $T_K=129$ MeV, as extracted in Ref.~~\cite{STAR:2017sal} from the transverse momentum spectra of identified particles using the blast-wave model.  Following Ref.~\cite{Sung:2024vyc}, the time evolution of the temperature, volume, and pion fugacity of the hadronic matter is determined by imposing four constraints: the conservation of the effective pion, kaon, and nucleon numbers, and constant entropy per hadron during the evolution~\cite{Xu:2017akx}. These conditions yields a hadronic volume of $V_K=403$ fm$^3$ and a pion fugacity $z_\pi=1.11$ at kinetic freeze-out.  

With the chemical and kinetic freeze-out times obtained from the AMPT model simulations of these collisions~\cite{Xu:2017akx}, namely $\tau_{C}=5.13$ fm/$c$ and $\tau_K=6.46$ fm/$c$, the time evolution of the temperature and volume of the hadronic matter can be parametrized as follows:
\begin{eqnarray}
&&T(\tau) = T_{C} -(T_{C}-T_K)\left(\frac{\tau-\tau_C}{\tau_K-\tau_C}\right)^{0.9},\notag\\
&&V(\tau) =V_C+(V_K-V_C)\left(\frac{\tau-\tau_C}{\tau_K-\tau_C}\right).\notag\\
\end{eqnarray}
Because the pion chemical potential increases only slightly as the temperature decreases, we neglect its temperature dependence 
and take its value to be constant and equal to one in the present study.

\begin{figure}[h]
    \centering
    \includegraphics[width=0.8\linewidth]{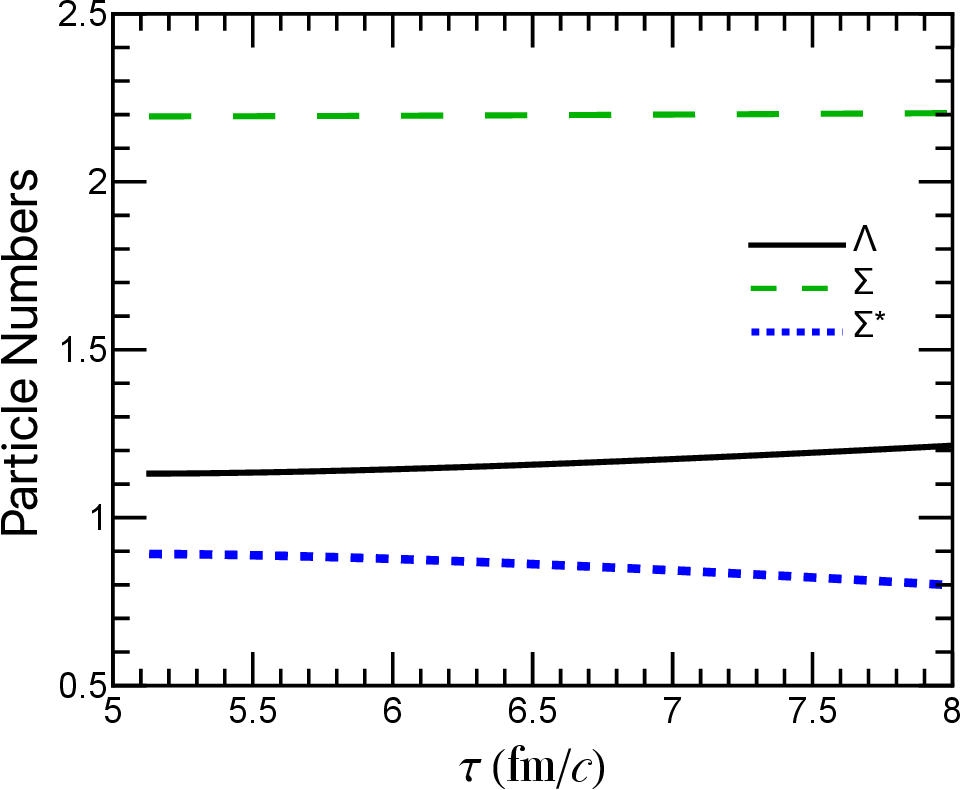}
    \caption{Time evolution of $\Lambda$ (solid line), $\Sigma$ (dashed line), and $\Sigma^*$ (dotted line) numbers in Au-Au collisions at $\sqrt{s_{NN}}=7.7$ GeV and 20-50\% collision centrality. } 
     \label{fig:numbers}
\end{figure}

For the initial $\Lambda$, $\Sigma$, and $\Sigma^*$ numbers at $T_C=146$ MeV, we take their values from the statistical model, i.e., $N_Y^{(0)}=N_Y^{(T_C)}$ and $N_{\Sigma^*}^{(0)}=N_{\Sigma^*}^{(T_C)}$, where $N_Y^{(T_C)}$ and $N_{\Sigma^*}^{(T_C)}$ are the equilibrium numbers of $\Lambda$ and $\Sigma^*$ at $T_C$. The results from solving the first two equations of Eq.(\ref{kinetic3}) are shown in Fig.~\ref{fig:numbers}, which displays the time evolution of $\Lambda$ number (solid line), $\Sigma$ number (dashed line), and $\Sigma^*$ number (dotted line). Both $\Lambda$ and $\Sigma$ numbers increase slightly with time, while the $\Sigma^*$ number decreases at a similar rate, consistent with the conservation of their total number. 

\begin{figure}[h]
    \centering
    \qquad\includegraphics[width=0.8\linewidth]{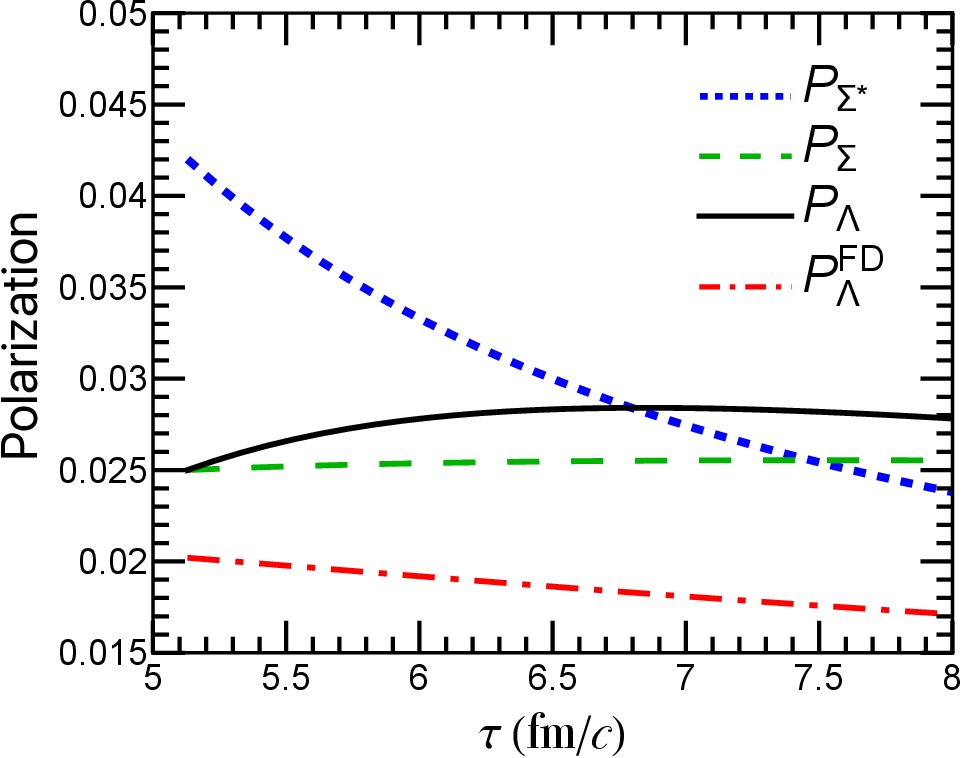}   
    \caption{Time evolution of $\Lambda$ (solid line), $\Sigma$ (dashed line), and $\Sigma^*(1385)$ (dotted line) polarizations in Au-Au collisions at $\sqrt{s_{NN}}=7.7$ GeV and 20-50\% collision centrality, for initial $\Lambda$ and $\Sigma$ polarizations of 0.025 and initial $\Sigma^*$ polarization of 0.042. Also shown (dot-dashed line) is the $\Lambda$ polarization $P_\Lambda^{\rm FD}$ that includes the feed-down contributions described in the text.}
     \label{fig:polarization}
\end{figure}

For the time evolution of the $\Lambda$ polarization, we take its initial value to be $P_\Lambda^{(0)}=0.025$, as obtained in Ref.~\cite{Li:2017slc} using the temperature and thermal vorticity calculated from a multiphase transport (AMPT) model~\cite{Lin:2004en} via the course-grained method for Au-Au Collisions at $\sqrt{s_{NN}}=7.7$ GeV and averaged over two impact parameters of 7 fm and 9 fm.  This value is close to the upper limit of the $\Lambda$ polarization measured by the STAR Collaboration for 20-50\% collision centrality~\cite{STAR:2017ckg}.   For the initial polarizations of $\Sigma$ and $\Sigma^*$, we assume they are in thermal equilibrium with the same vorticity at $T_C$, as in the case for the $\Lambda$ polarization.  Using the non-relativistic expression for vorticity $\mathbf\omega=\frac{1}{2}\nabla\times{\bf v}$, where ${\bf v}$ is the velocity field, the polarization of a spin 1/2 hadron is approximately $\frac{\omega}{2T_C}$, while that of a spin 3/2 hadron is approximately $\frac{5\omega}{6T_C}$ for $\omega/T_C\ll 1$~\cite{Becattini:2016gvu}.  We therefore take $P_\Sigma^{(0)}=P_\Lambda^{(0)}=0.025$ and $P_{\Sigma^*}^{(0)}=(5/3)P_\Lambda^{(0)}=0.042$.  

The time evolution of $\Lambda$, $\Sigma$, and $\Sigma^*$ polarizations, obtained by solving the last two equations of Eq.(\ref{kinetic3}) with the time-dependent $\Lambda$, $\Sigma$ and $\Sigma^*$ numbers from Fig.~\ref{fig:numbers}, is shown in Fig.~\ref{fig:polarization}.    The $\Lambda$ polarization increases slightly with time and reaches about 0.028 at $T_K$ when the $\Sigma^*$ polarization decreases from its initial value of 0.042 to 0.03 at $T_K$.  Compared with the $\sim$7\% decrease in $\Lambda$ polarization found in Ref.~\cite{Sung:2024vyc} when the $\Sigma^*$ resonance was not treated as a dynamic degree of freedom, the present results show a slightly enhancement of $\Lambda$ polarization due to hadronic scattering effects during the hadronic stage of relativistic heavy ion collisions.  If the hadronic stage were to last twice as long, ending at $\tau=8 ~{\rm fm}/c$ when the temperature has dropped to $T=112$ MeV, the $\Lambda$ polarization remains to have essentially the same value of 0.028.  The $\Sigma^*$ polarization, on the other hand, continuously decreases with time to the value 0.024.  In contrast, the time evolution of the $\Sigma$ polarization is minimal due to the smaller $\Sigma+\pi\to\Sigma^*$ cross section compared to that of $\Lambda+\pi\to\Sigma^*$.  

\section{Feed-down contributions from $\Xi$ and $\Xi^*$ to $\Lambda$ polarization}\label{feed-down}

\begin{figure}[h]
    \centering  \includegraphics[width=0.9\linewidth]{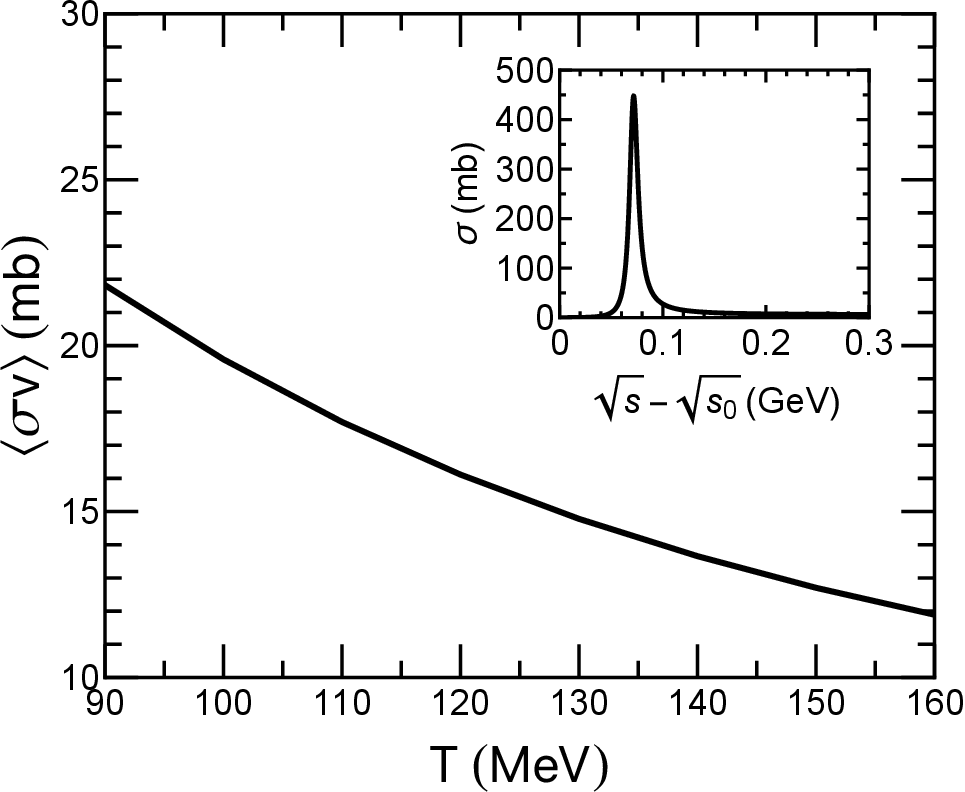}
     \caption{Temperature dependence of the thermal average of the spin-averaged $\Xi+\pi\rightarrow \Xi^*$ cross section shown in inset.}
     \label{fig:thermal1}
\end{figure}

Since the $\Lambda$ polarization measured in experiments includes not only the primarily produced $\Lambda$ hyperons but also those from the decays of other hadrons~\cite{Becattini:2016gvu,Xia:2019fjf} -- particularly from the strong decay $\Sigma^*\to\Lambda+\pi$, the electromagnetic decay $\Sigma^0\to\Lambda+\gamma$, and the weak decays $\Xi(1315)\to\Lambda+\pi^0$ and $\Xi^-(1322)\to\Lambda+\pi^-$ -- it is important to account for these feed-down contributions.  For the $\Xi$ hyperon, its scattering with pions is dominated by $\Xi^*(1532)$ resonance, with a spin-averaged cross section of a similar form to that in Eq.(\ref{eq:resonance}) for $\Lambda-\pi$ and $\Sigma-\pi$ scatterings via the $\Sigma^*$ resonance. Using the spins of $s_\Xi=1/2$ and $s_{\Xi^*}=3/2$ and the decay width $\Gamma_{\Xi^*}(m_{\Xi^*})=10$ MeV, the $\Xi-\pi$ scattering cross section and its thermal average are shown by the solid lines in Fig.~\ref{fig:thermal1}.  Analogous to the kinetic equations in Eq.(\ref{kinetic3}) for $\Lambda$, $\Sigma$, and $\Sigma^*$ numbers and polarizations, the time evolution of $\Xi$ and $\Xi^*$ numbers and polarizations in relativistic heavy ion collisions is given by 
\begin{eqnarray}\label{eq:rate3}
&&\frac{dN_{\Xi}}{d\tau} = \langle\sigma_{\Xi+\pi\to\Xi^*} v\rangle z_{\pi}^{(T)}n_{\pi}^{(T)}\left(\frac{N_\Xi^{(T)}}{z_\pi^{(T)}N_{\Xi^*}^{(T)}} N_{\Xi^*}-N_\Xi\right),\notag\\   
&&\frac{dN_{\Xi^{*}}}{d\tau}=\langle \sigma_{\Xi+\pi\to\Xi^*} v\rangle z_{\pi}^{(T)}n_{\pi}^{(T)}\notag\\
&&\hspace{1.3cm}\times\left(N_\Xi-\frac{N_\Xi^{(T)}}{z_\pi^{(T)}N_{\Xi^*}^{(T)}}N_{\Xi^*}\right),\notag\\
&&\frac{dP_\Xi}{d\tau}=\langle\sigma_{\Xi+\pi\to\Xi^*} v\rangle n_{\pi}^{(T)}\frac{N_\Xi^{(T)}N_{\Xi^*}}{N_{\Xi^*}^{(T)}N_\Xi}(P_{\Xi^*}-P_{\Xi}),\notag\\
&&\frac{dP_{\Xi^*}}{d\tau}=\langle\sigma_{\Xi+\pi\to\Xi^*} v\rangle z_{\pi}^{(T)}n_{\pi}^{(T)}\frac{N_\Xi}{N_\Xi^*}\left(\frac{5}{9}P_\Xi-P_{\Xi^*}\right).
\end{eqnarray}
As in the case of $\Lambda$, $\Sigma$ and $\Sigma^*$, the total number of $\Xi$ and $\Xi^*$ remains constant during the hadronic evolution.  The $\Xi$ polarization increases or decreases with time depending on whether $P_{\Xi^*}-P_\Xi$ is  positive or negative, while  the time dependence of the $\Xi^*$ polarization is proportional to $5P_\Xi/9-P_{\Xi^*}$.

\begin{figure}[h]
    \centering
    \includegraphics[width=0.8\linewidth]{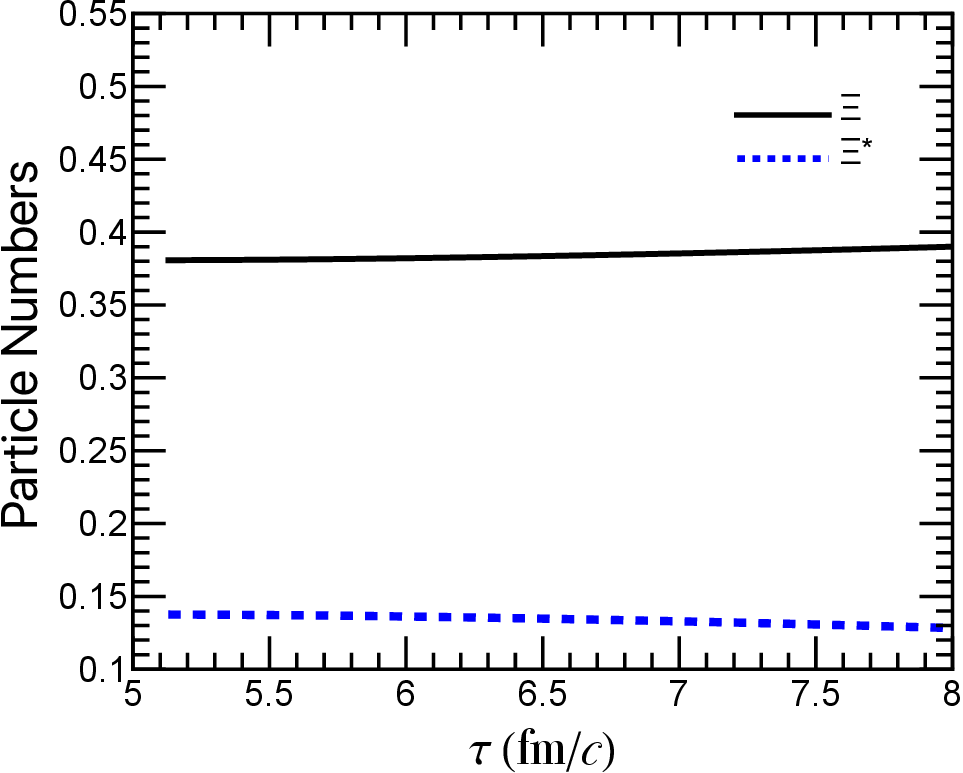}
    \caption{Time evolution of the $\Xi$ number (solid line) and $\Xi^*$ number (dotted line) in Au-Au collisions at $\sqrt{s_{NN}}=7.7$ GeV and 20-50\% collision centrality. } 
     \label{fig:numbers1}
\end{figure}

\begin{figure}[h]
    \centering
    \qquad\includegraphics[width=0.8\linewidth]{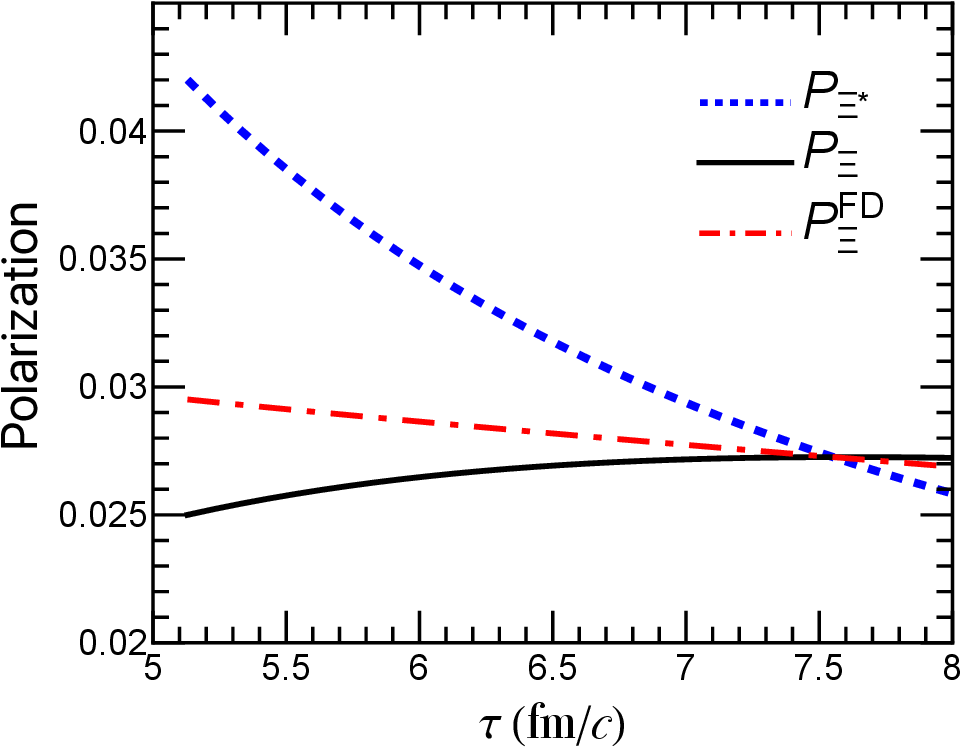}   
    \caption{Time evolution of $\Xi$ (solid line) and $\Xi^*$ (dotted line) polarizations in Au-Au collisions at $\sqrt{s_{NN}}=7.7$ GeV and 20-50\% collision centrality, with initial $\Xi$ polarizations of 0.025 and $\Xi^*$ polarization of 0.042. The dot-dashed line denotes the $\Xi$ polarization $P_\Xi^{\rm FD}$ that includes the feed-down contribution from $\Xi^*$.} 
     \label{fig:polarization1}
\end{figure}

Using the initial $\Xi$ and $\Xi^*$ numbers from the statistical model at $T_C$, calculated from similar expressions as in Eq.(\ref{equilibrium}) with corresponding masses, width, and normalization constant $A_{\Xi^*}=0.63$, and initial $\Xi$ polarization of 0.025 and $\Xi^*$ polarization of 0.042 -- values taken to match those for $\Sigma$ and $\Sigma^*$ hyperons -- the results from solving Eq.(\ref{eq:rate3}) for the time evolution of $\Xi$ and $\Xi^*$ numbers and polarizations are shown in Fig.~\ref{fig:numbers1} and Fig.~\ref{fig:polarization1}, respectively.  Similar to $\Lambda$, $\Sigma$, and $\Sigma^*$, the $\Xi$ number increases with time while the $\Xi^*$ number decrease, keeping their sum constant.  Correspondingly,  the $\Xi$ polarization increases over time, while the $\Xi^*$ polarization decreases.

With the time dependence of $\Lambda$, $\Sigma$, and $\Xi$ polarizations, the time dependence of the $\Lambda$ polarization -- including feed-down contributions from the decays $\Sigma^0\to\Lambda+\gamma$, $\Sigma^*\to\Lambda+\pi $, and $\Xi\to\Lambda+\pi$ -- can then be calculated as follows:
\begin{eqnarray}\label{eq:FD}
P_\Lambda^{\rm FD}=\frac{\sum_{Y=\Lambda,\Sigma^0,\Sigma^{*+},\Sigma^{*0},\Sigma^{*-},\Xi^0,\Xi^-}C_YP_YN_Y}{\sum_{Y=\Lambda,\Sigma^0,\Sigma^{*+},\Sigma^{*0},\Sigma^{*-},\Xi^0,\Xi^-}N_Y},
\end{eqnarray}
where the $C_Y$ are the polarization transfer coefficients of hyperons to the $\Lambda$ in their decay. Their values for $\Lambda$, $\Sigma^0$, $\Xi^0$, and $\Xi^-$ are $C_\Lambda=1$, $C_{\Sigma^0}=-1/3$, $C_{\Xi^0}=0.900$, and $C_{\Xi^{-}}=0.927$~\cite{Becattini:2016gvu,Xia:2019fjf}.  For $\Sigma^{^+}$, $\Sigma^{*0}$, and $\Sigma^{*-}$, their polarization transfer coefficients are determined as follows. The $\Sigma^{*0}$ has the strong decay channels $\Sigma^{*0}\to\Lambda+\pi^0$, $\Sigma^{*0}\to\Sigma^+ +\pi^-$, and $\Sigma^{*0}\to\Sigma^- +\pi^+$ but only the first one contributes to the $\Lambda$ polarization, with $C_{\Sigma^{*0}}=0.88$. For $\Sigma^{*\pm}$, the strong decays include $\Sigma^{*\pm}\to\Lambda+\pi^\pm$, $\Sigma^{*\pm}\to\Sigma^\pm+\pi^0$, and $\Sigma^{*\pm}\to\Sigma^0+\pi^\pm$. Taking into account the branching ratios of 88\% and 12\% for $\Sigma^*$ decaying to $\Lambda$ and $\Sigma$, respectively, and the equal decay probabilities to charged and neutral $\Sigma$, we obtain the polarization transfer coefficients $C_{\Sigma^{*+}}=C_{\Sigma^{*-}}=0.88+0.12\times\frac{1}{2}\times(-\frac{1}{3})=0.86$, since only $\Sigma^0$ can decay electromagnetically to $\Lambda+\gamma$, with a polarization transfer coefficient of $-1/3$.

Using the time dependence of $\Lambda$, $\Sigma$, and $\Sigma^*$ numbers in Fig.~\ref{fig:numbers} and of the $\Xi$ number in Fig.~\ref{fig:numbers1}, which give $N_{\Sigma^0}=N_\Sigma/3$ and $N_{\Xi^0}=N_{\Xi^-}=N_\Xi/2$, as well as the time dependence of their polarizations from Figs.~\ref{fig:polarization} and \ref{fig:polarization1}, the resulting time dependence of $P_\Lambda^{\rm FD}$ is shown by the dot-dashed line in Fig.~\ref{fig:polarization}.  It is seen that $P_\Lambda^{\rm FD}$ has an initial value of about 0.02 at $\tau_C=5.13$ fm/$c$, decreases to 0.019 at $\tau_K=6.46$ fm/$c$, and becomes 0.017 at $\tau=8$ fm/$c$, which are smaller than the value of $P_\Lambda$ by 20\%, 34\%, and 39\% at corresponding times.  

The reduction of the $\Lambda$ polarization at $T_C$ due to the feed-down contribution in our study has a similar magnitude to that found in Refs.~\cite{Li:2017slc,Becattini:2019ntv}.  This reduction would have been much larger if the positive feed-down contribution from the strong decay of $\Sigma^*$ were not included, because of the large negative feed-down from the electromagnetic decay of $\Sigma^0$ and the small positive contributions from the weak decays of $\Xi^0$ and $\Xi^-$ to $\Lambda$, due to their relatively small numbers resulting from their large masses. 

Results from the present study, which is more realistic than Ref.~\cite{Sung:2024vyc}, thus indicate the polarization of directly produced $\Lambda$ increase slightly during the hadronic stage of relativistic heavy ion collision, which is in contrary to the continuously slow decrease with time found in Ref.~\cite{Sung:2024vyc}, but that including the feed-down contribution decreases over time.  However, these changes in the $\Lambda$ polarizations with and without feed-down contributions during the hadronic stage are only about 10-15\%, indicating that the spin degree of freedom of a $\Lambda$ essentially decouples at the chemical freeze out of relativistic heavy ion collisions when it is produced from the quark-gluon plasma. 

\section{conclusions}\label{conclusions}

We have studied the effect of the $\Sigma^*$ resonance on $\Lambda$ polarization in relativistic heavy ion collisions by including its production from $\Lambda+\pi\to\Sigma^*$ and $\Sigma+\pi\to\Sigma^*$ scatterings, as well as its decay through the inverse processes $\Sigma^*\to\Lambda+\pi$ and $\Sigma^*\to\Sigma+\pi$, in the spin-dependent kinetic equations. By converting these equations into coupled kinetic equations for the numbers and polarizations of $\Lambda$, $\Sigma$, and $\Sigma^*$, we are able to study the time evolution of their abundances and polarizations in Au-Au collisions at $\sqrt{s_{NN}}=7.7$ GeV and 20-50\% collision centrality, as measured at RHIC.

In contrast to the slight decrease of $\Lambda$ polarization with time found in Ref.~\cite{Sung:2024vyc}, where the $\Sigma^*$ resonance was not explicitly included as a dynamic degree of freedom, we find that the $\Lambda$ polarization increases with time when the $\Sigma^*$ polarization is larger than the $\Lambda$ polarization, as determined by their respective equilibrium values in the thermal vorticity at chemical freeze-out.  As the $\Sigma^*$ polarization gradually decreases with time, the growth of the $\Lambda$ polarization slows down, eventually reaching a final value higher than its initial value at $T_C$.  

However, when the feed-down contributions to the $\Lambda$ polarization from the electromagnetic decay of $\Sigma$, the strong decay of $\Sigma^*$, and the weak decay of $\Xi$ hyperons are included -- using their polarizations obtained from the solutions of their coupled kinetic equations -- the resulting $\Lambda$ polarization is reduced and shows a decreasing trend during the expansion and cooling of the hadronic matter.  These changes in $\Lambda$ polarization, with and without feed-down contributions, are nevertheless modest, supporting the approximation that the spin degree of freedom of a $\Lambda$ essentially freezes out at the chemical freeze-out of relativistic heavy ion collisions.  This validates the common practice of comparing theoretical results with experimental data under the assumption of the decoupling of $\Lambda$ spin at the chemical freeze-out of heavy ion collisions.  

The above conclusion is also true for the $\Xi$ polarization because it only increases by less than 10\% during the hadronic stage as shown by the solid line in Fig.~\ref{fig:polarization1}.  Including the feed-down contribution from the $\Xi^*$ due to the decay $\Xi^*\to\Xi+\pi$ does not change this conclusion as shown by the dot-dashed line in the same figure. The latter, denoted by $P_\Xi^{\rm FD}$, is calculated according to 
\begin{eqnarray}
P_\Xi^{\rm FD}=\frac{P_\Xi N_\Xi+P_{\Xi^*}N_{\Xi^*}}{N_\Xi+N_{\Xi^*}},
\end{eqnarray}
where a value of one is used for the polarization transfer coefficient of $\Xi^*$ to $\Xi$~\cite{Xia:2019fjf}.   Experimentally, the STAR Collaboration has measured the $\Xi$ polarization in Au+Au collisions at the collision energy of $\sqrt{s_{NN}}=200$ GeV and centrality of 0-50\% with the value of 0.47$\pm$0.10(stat)$\pm$0.23(syst)\%~\cite{STAR:2020xbm}. The value of $\Xi$ polarization is shown to increase with decreasing collision energy, reaching a value of about 1.5\% at $\sqrt{s_{NN}}=7.7$ GeV according to the calculations in Ref.~\cite{Wei:2018zfb}, which assumes equilibrium between the $\Xi$ spin and the thermal vorticity from the course-grained a multiphase transport (AMPT) model~\cite{Lin:2004en} at the hadronization of produced quark-gluon plasma.  This value of the $\Xi$ polarization is smaller than our value of about 2.7\% with or without the feed-down contribution as a result of our use of the non-relativistic approximation for the determination of initial $\Xi$ and $\Xi^*$ polarizations, which neglects the mass ordering of hadron polarizations by the thermal vorticity.  

Our spin-dependent kinetic approach can further be used to study the hadronic effects on the polarization of $\Omega(1672)$ hyperons, e.g.  by considering their scatterings with pions and kaons through its resonances~\cite{ParticleDataGroup:2024cfk}. Such a study will allow us to check if the $\Omega(1672)$ polarization also freezes out earlier as assumed in Ref.~\cite{Wei:2018zfb}, which is essential for making meaningful comparisons with the measurements by the STAR Collaboration~\cite{STAR:2020xbm}.

\section*{ACKNOWLEDGEMENTS}

This work was supported by Academia Sinica under Project No. AS-CDA-114-M01 (H.S.); the U.S. Department of Energy under Award No. DE-SC0015266 (C.M.K.); and the National Research Foundation of Korea under Grant Nos. 2023R1A2C300302311 and 2023K2A9A1A0609492411 (S.H.L.).

\bibliography{main.bib}
\end{document}